\documentclass[twocolumn]{aa}

\usepackage{amssymb,amsmath}
\usepackage{mhchem}
\usepackage{threeparttable, tablefootnote}

\usepackage{graphicx}
\usepackage{txfonts}
\begin{document}

\title{A major asymmetric ice trap in a planet-forming disk:}
\subtitle{II. prominent \ce{SO} and \ce{SO_2} pointing to C/O$<$1}

\titlerunning{ \ce{SO} and \ce{SO_2} in the IRS~48 disk}
\authorrunning{A. S. Booth et al.}

\author{Alice S. Booth \inst{1}, Nienke van der Marel \inst{2,3}, Margot Leemker \inst{1}, Ewine F. van Dishoeck \inst{1,4}, Satoshi Ohashi\inst{5}}

\institute{Leiden Observatory, 
Leiden University, 
2300 RA Leiden, 
the Netherlands \\
\email{abooth@strw.leidenuniv.nl} 
\and
Physics \& Astronomy Department, 
University of Victoria, 
3800 Finnerty Road, 
Victoria, BC, V8P 5C2, 
Canada 
\and
Banting Research fellow
\and
Max-Planck-Institut f\"{u}r Extraterrestrishe Physik, Gie{\ss}enbachstrasse 1, 85748 Garching, Germany
\and
RIKEN Cluster for Pioneering Research, 2-1, Hirosawa, Wako-shi, Saitama 351-0198, Japan
}

\date{Received 12/04/2021; Accepted 7/06/2021}

 
  \abstract{
Gas-phase sulphur bearing volatiles appear to be severely depleted in protoplanetary disks. The detection of CS and non-detections of SO and \ce{SO_2} in many disks have shown that the gas in the warm molecular layer, where giant planets accrete their atmospheres, has a high C/O ratio. In this letter, we report the detection of SO and \ce{SO_2} in the Oph-IRS~48 disk using ALMA. This is the first case of prominent \ce{SO_2} emission detected from a protoplanetary disk. The molecular emissions of both molecules is spatially correlated with the asymmetric dust trap. We propose that this is due to the sublimation of ices at the edge of the dust cavity and that the bulk of the ice reservoir is coincident with the millimeter dust grains. Depending on the partition of elemental sulphur between refractory and volatile materials the observed molecules can account for 15-100\% of the total volatile sulphur budget in the disk. In strong contrast to previous results, we constrain the C/O ratio from the CS/SO ratio to be $<$ 1 and potentially solar. This has important implications for the elemental composition of planets forming within the cavities of warm transition disks.}


   \maketitle
%

\section{Introduction} 

New planetary systems are made from dust and gas in the rotating
disks around young stars. Elemental abundances and ratios such as
C/O are key quantities in linking planet composition to their
formation history \citep{_berg_2011, 2017MNRAS.469.3994B, 2018A&A...613A..14E}. Sulphur is a particularly interesting element for astrophysics, vulcanology on young planets, and
formation of life but its main reservoirs remain a puzzle.
The depletion of elemental sulphur by 1-3 orders of magnitude in regions of dense gas has been an outstanding problem in astrochemistry for decades \citep[e.g.][]{1994A&A...289..579T, Dutrey1997, 2009ApJ...700.1299J}. The bulk of the sulphur is thought to reside in refractory materials but the exact reservoir is still unconfirmed \citep{2002Natur.417..148K, 2015MNRAS.450.1256W}. 
In dark clouds and young ($<$1~Myr) Class 0/I sources sulphur-bearing molecules are routinely detected in the gas-phase but their contribution to the total sulphur budget and chemical origin are still somewhat unclear \citep[e.g.][]{2019A&A...626A..71A, 2020ApJ...898..131L, 2020arXiv201207667G}.  
On the ices, less is known aside from the direct detections of \ce{SO_2} and \ce{OCS}, and an upper-limit for \ce{H_2S}, contributing to $\approx$5\% of the total sulphur budget in protostars \citep{1997A&A...317..929B,2015ARA&A..53..541B}. Indirectly, the ice composition can be inferred from observations of environments where ices sublimate, e.g., hot cores, and the gas detected in these systems suggests the presence of \ce{SO} and \ce{SO_2} ices \citep[][]{2018MNRAS.476.4949D, Tychoniec2021}.

In protoplanetary disks, a study of the elemental composition of the accreted disk material in the photopsheres of Herbig Ae/Be stars shows that $\approx$89$\pm$8$\%$ of the sulphur in disks is refractory \citep{2019ApJ...885..114K}. Therefore, the quest has been ongoing to detect different molecules to try and account for the total volatile fraction of sulphur in $>$1~Myr protoplanetary disks especially as this is where gas-giant planets accrete the bulk of their atmospheres. 
Sulphur species may form of hazes in the atmosphere \citep{2020NatAs...4..986H} and they are important for pre-biotic chemistry \citep{2018AsBio..18.1023R}.
\ce{CS} is most commonly detected followed by \ce{H_2CS} \citep[][]{2019ApJ...876...72L, 2020arXiv201102305C,2020A&A...638A.110F}, yet although CS is the most abundant sulphur species in protoplanetary disks, it only accounts for $<$1\% of the total sulphur budget. 
There are only a few sources with detections of SO including the two Herbig disks, AB~Aur and HD~100546 \citep{2016A&A...592A.124G, 2010A&A...524A..19F, 2018A&A...611A..16B}. There are one tentative detection of \ce{SO_2} \citep[][]{2018A&A...617A..28S} and one detection of \ce{H_2S} \citep{2018A&A...616L...5P}. 

Even-though only a small fraction of volatile sulphur has been detected in disks so far, sulphur-species have provided key insight into the elemental make up of disk's atmospheres.
The CS/SO ratio has been proposed as a tracer of the underlying C/O ratio in the gas.
Detections of CS and non-detections/tentative detections of \ce{SO} and \ce{SO_2} in many disks (LkCa~15, DM~Tau, GO~Tau, MWC~480 and PDS~70) indicate a high gas phase C/O ratio around unity in the warm molecular layer \citep[][]{2011A&A...535A.104D, 2018A&A...617A..28S, 2021arXiv210108369F}. These observations are directly probing where gas giant planets accrete their atmospheres \citep{c8f807d3c6ee4380a6de362147fe4c35,2019Natur.574..378T,2020A&A...635A..68C}. This high ratio is in line with the depletion of volatile C and O in disks that can be explained by the vertical mixing, radial transport, freeze-out and subsequent chemical reprocessing of CO \citep[e.g.][]{2020ApJ...899..134K}, where more O than C is locked up in the ices. Observing sulphur-species is therefore an extremely useful and complementary tool in constraining the bulk elemental composition of the gas in disks. 

Here we present the first measurements of \ce{SO} and \ce{SO_2} in a disk that can be directly linked to an ice trap. 
Oph-IRS 48 is a 2.0~M$_{\odot}$ Herbig A0 source located at 134~pc in the $\rho$~Ophiuchus star forming region \citep{2012ApJ...744..116B, Gaia2018}. This protostar is host to a transition disk with an extreme azimuthal asymmetry traced in the millimeter to cm sized grains \citep{vanderMarel2013, vanderMarel2015vla} whereas the micron sized grains are distributed axisymmetrically \citep{2007A&A...469L..35G}. The dust cavity extends out to 60~au and the gas cavity traced in CO to 25~au with a depletion of gas in the cavity of $\times 10^3$ relative to the ring \citep{vanderMarel2016-isot}. The disk is gas rich and warm with detections of the rarer CO isotopologues \ce{C^{18}O} and \ce{C^{17}O}, and \ce{H_2CO} \citep{Bruderer2014, vanderMarel2014, vanderMarel2016-isot}. The bulk of the gas, traced in CO, is present through-out the entire azimuth of disk, like the micron-sized grains, whereas the \ce{H_2CO} was tentatively found to be co-spatial with the dust trap. This source presents a unique opportunity to study the chemistry associated with an exposed inner cavity edge and a high concentration of large icy dust grains.
In \citet{vanderMarel2021}, hereafter Paper I,  we present new detections of \ce{H_2CO} and \ce{CH_3OH} in the Oph-IRS~48 protoplanetary disk and here in Paper II we present observations of the S-bearing molecules: \ce{SO_2}, \ce{^{34}SO_2}, \ce{SO} and \ce{CS}. In Section 2 we outline these observations and in Section 3 we show the resulting images and inferred column densities. In Section 4 we discuss the total volatile sulphur detected in the disk compared to the total Sulphur budget, investigate the chemical origin of the detected species and the link to the dust trap, and provide key constraints on the C/O of the gas that differ from other disks observed so far.  
 
\begin{center}
\begin{table*}
\caption{Properties of the molecular lines and images analysed in this work.}
\small
\centering
\begin{tabular}{clcccccccc}
\hline \hline
Molecule & Transition & Frequency & $\mathrm{log_{10}(\textit{A}_{ij})}$ & ${E_{\mathrm{up}}}$   & $g_{\mathrm{u}}$    &  Beam  &  $\Delta$v  & Peak~~~rms$^*$ & Disk Int. Flux \\  
         &            &  (GHz)    & ($\mathrm{s^{-1}}$) & (K)              &   &  & km$\mathrm{s^{-1}}$ & (mJy beam$^{^{-1}}$)&  (mJy km s$^{^{-1}}$) \\ 
\hline 
\ce{C^{18}O}  & 3 - 2    & 329.3305525  & -5.6631  &31.6&7 & 0\farcs20$\times$0\farcs16~(65$^{\circ}$) &0.23& 58.45~~~5.10  & 896.5 $\pm$ 90 \\
\ce{SO_2}   &   \ce{5_{3,3}} - \ce{4_{2,2}}      & 351.2572233 & 	-3.0672  &35.9 & 11 &0\farcs55$\times$0\farcs44~(80$^{\circ}$) & 1.7 &34.33~~~0.91  & 305.1 $\pm$ 31\\
\ce{SO_2}   &   \ce{14_{4,10}} - \ce{14_{3,11}}      &  351.8738732 &  -3.4644 & 135.9 & 29  &0\farcs55$\times$0\farcs44~(80$^{\circ}$) & 1.7 &13.87~~~0.91 & 124.9 $\pm$ 12 \\
\ce{^{34}SO_2} &   \ce{6_{3,3}} - \ce{5_{2,4}}   & 362.1582327 & -3.4839  & 40.7  &  13&0\farcs53$\times$0\farcs42~(80$^{\circ}$) &1.7&12.11~~~1.20  & 87.3 $\pm$ 9
\\
\ce{SO}  &   \ce{1_{2}} - \ce{0_{1}}      &  	329.3854770 & -4.8467& 15.8  & 3 & 0\farcs20$\times$0\farcs16~(65$^{\circ}$) & 1.7$^{**}$& 13.62~~~2.49  & 166.9 $\pm$ 17\\
\ce{CS}  &   \ce{7-6}      &  342.8828503 	 &	-3.0774	& 65.8 & 15 & 0\farcs20$\times$0\farcs16~(65$^{\circ}$) &0.23& ~~ ~ -~~~~~~~8.23  & $<$348 $\pm$ 35\\
\hline
\end{tabular}
\label{line_list}
\begin{tablenotes}\footnotesize
\item{$^*$ per channel}
\item{$^{**}$ Note the native channel width of these data is 0.23  km s$^{^{-1}}$.}
\item{The values for the line frequencies, Einstein A coefficients, and upper energy levels ($E_{up}$) and degeneracies ($\mathrm{g_u}$) are taken from the Cologne Database for Molecular Spectroscopy (CDMS) \citep{2005JMoSt.742..215M}.}
\end{tablenotes}
\end{table*}  
\end{center}

\section{Observations}

We utilise two sets of Band 7 observations of Oph-IRS~48 taken with the Atacama Large Millimeter/submillimeter Array (ALMA) during Cycles 2 and 5. 
The Cycle 2 program 2013.1.00100.S (PI: Nienke van der Marel) were taken June and August 2015 and were calibated using using the provided calibration scripts.
These data cover one transition of \ce{C^{18}O}, \ce{SO} and \ce{CS}  (see Table~\ref{line_list}) 
and have native channel width of 122~kHz or $\approx$0.23~km s$^{-1}$ and a spatial resolution of $\approx$0\farcs2. 
The Cycle 5 program 2017.1.00834.S (PI: Adriana Pohl) was taken in August 2018. For full details on the data calibration see \citet{Ohashi2020} where the continuum polarization data are  presented.  This paper focuses on the detection of two \ce{SO_2} lines and one \ce{^{34}SO_2} line (see Table~\ref{line_list}) in these data. The detected \ce{H_2CO} and \ce{CH_3OH} lines are the focus of Paper I. These line data all have a channel width of 1953 kHz or $\approx$1.7 km s$^{-1}$ and a spatial resolution of $\approx$0\farcs5. 

We use the continuum images from both 2013.1.00100.S and \citet{Ohashi2020}, and the higher spatial resolution \citet{Francis2020} image to compare to the respective line images. All of the lines were imaged using CASA \texttt{tCLEAN} with robust weighting of 0.5 and a Keplerian mask (with an inclination angle of 50$^{\circ}$ and position angle of 100$^{\circ}$, \citealt[e.g.][]{2021AJ....161...33V}) and stellar mass of 2.0~M$_{\odot}$ and distance of 134~pc. Using the phase-centre variable in \texttt{tCLEAN} the  central position is set to ICRS 16h27m37.1797s, -24$^{\circ}$30'35.480" computed using the Gaia DR2 position at the time of the observations \citep{Gaia2018} for the Cycle 5 data. 

The resulting beam sizes, peak emission and per channel rms are listed in Table~\ref{line_list}. All lines but the CS $J=7-6$ line are robustly detected ($>3\sigma$ emission over 3 consecutive channels) and the channel maps for the detected lines are shown in Appendix~\ref{fig:channelmaps}.
For the CS a 3$\sigma$ upper limit on the disk integrated flux (see Table~\ref{line_list}) was propagated from the rms noise in the  channel maps following the method in \citet{2019A&A...623A.124C}.
To reduce the per channel rms and thus increase the signal-to-noise the Cycle 2 SO \ce{1_{2}} - \ce{0_{1}} line was imaged at 1.7~km s$^{-1}$. This also matches the spectral resolution of the Cycle 5 \ce{SO_2} isotopologue lines.

\section{Analysis}

\begin{figure*}[h!]
    \centering
    \includegraphics[width=\hsize,trim={2cm 2cm 1cm 1cm},clip]{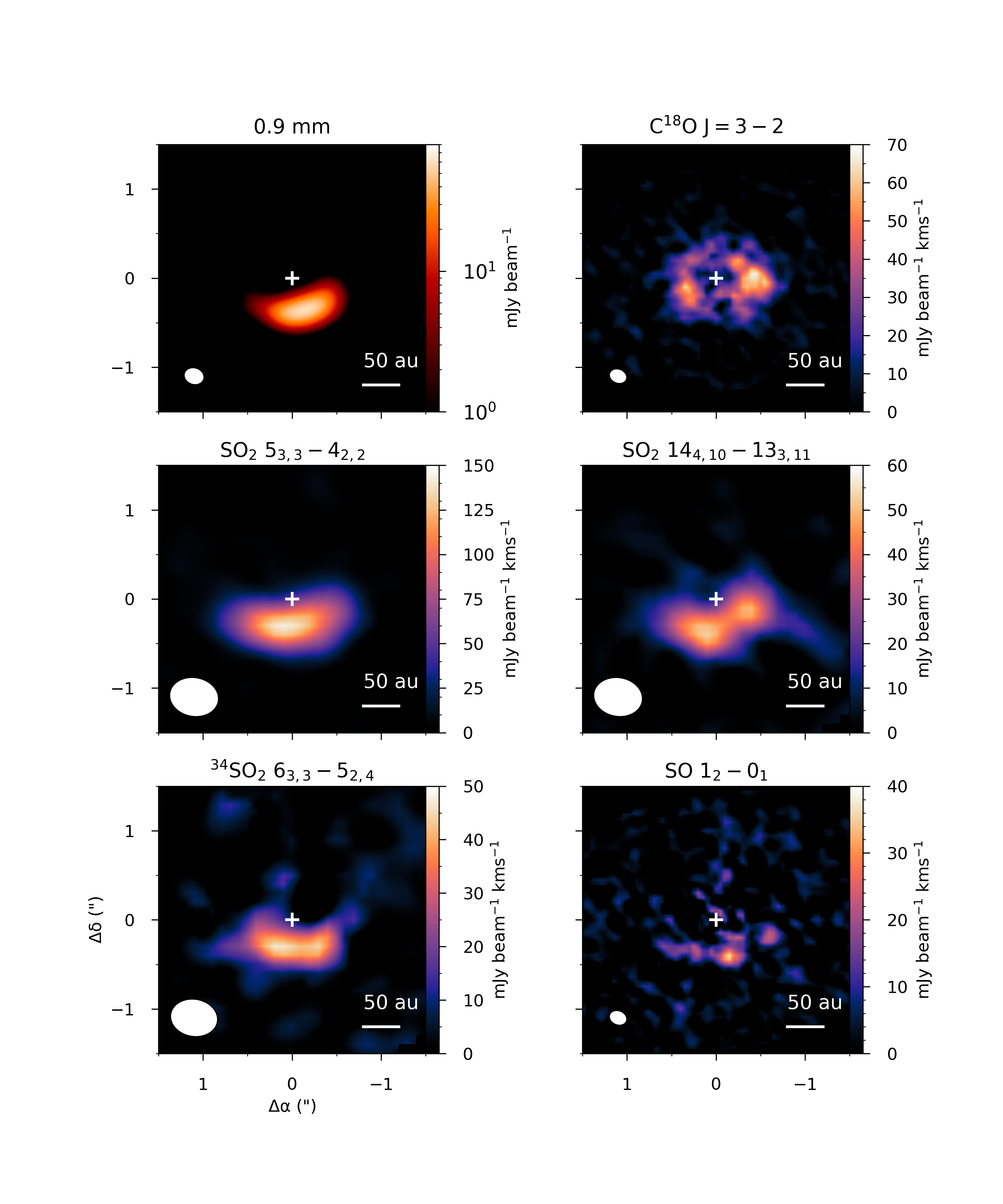}
    \caption{Integrated intensity maps of the 0.9~mm continuum emission and detected lines listed in Table~\ref{line_list}. The line maps are made with a Keplerian mask and the synthesised beam is shown in the bottom left corner of each panel.}
    \label{fig:mom0}
\end{figure*}

The Keplerian-masked integrated intensity maps for the \ce{C^{18}O}, \ce{SO_2}, \ce{^{34}SO_2} and \ce{SO} lines are presented in Figure~\ref{fig:mom0} alongside the $\approx$0\farcs2 0.9~mm continuum. 
\ce{SO_2} is clearly detected and this is the first prominent detection of this molecule in a protoplanetary disk. 
The \ce{SO_2}, \ce{^{34}SO_2} and \ce{SO} maps all show central dips in emission and are all highly asymmetric peaking in a similar azimuthal region of the disk as the mm-dust, much like the \ce{H_2CO} and \ce{CH_3OH} (see Paper I). 
This is in contrast to the \ce{C^{18}O} which does also show a central cavity but is present throughout the full azimuth of the disk. To further investigate the spatial relationship between the dust and sulphur-bearing molecules radial and azimuthal profiles of both were generated. The radial profiles for the dust and molecular lines were averaged over a wedge from 100$^{\circ}$ to 260$^{\circ}$ relative to the north minor axis and are shown Appendix Fig~\ref{fig:profiles}. 
The errors are given by the standard deviation of the intensity of the pixels (0\farcs1 or 0\farcs05 for the SO) in each radial bin 0\farcs2 for the \ce{SO_2} and \ce{^{34}SO_2} and 0\farcs1 for the SO) divided by the square root of the number of beams per arc of the wedge. This highlights the spread in intensity values per bin rather than the intrinsic errors in the data themselves. In the lower resolution data the \ce{SO_2} and \ce{^{34}SO_2} lines peak just outside the dust radial peak but the higher resolution \ce{SO} line is peaking at the same radial distance the dust emission. 

The azimuthal profiles were extracted from an ellipse with the same position and inclination angle of the disk. For the \ce{SO_2} and \ce{^{34}SO_2} a de-projected radius of $\approx$70~au was chosen and for \ce{SO} a radius of $\approx$90~au. The errors here are computed from the rms in the integrated intensity maps generated without a Keplerian mask. For comparison the normalised azimuthal profile of the 0.9~mm dust from the respective observations is shown. 
The \ce{SO_2} lines have a slightly wider azimuthal extent than the dust 
and the optically thinner lines have slight dips where the dust emission peaks, potentially indicating continuum absorption of the line emission. 
This may indicate that for the \ce{SO_2} the \ce{5_{3,3}} - \ce{4_{2,2}} line (${E_{\mathrm{up}}}$ of 36~K) is optically thick higher above the dust $\tau=1$ surface whereas the \ce{14_{4,10}} - \ce{14_{3,11}} line (${E_{\mathrm{up}}}$ of 136~K) is coming from a deeper region of the disk along the observing line of sight. 
In the Appendix in Fig~\ref{fig:comparison} we also show a comparison of the azimuthal profiles for the \ce{SO_2} with the \ce{H_2CO}, \ce{CH_3OH} from Paper I and the 0.9~mm dust. Here the \ce{H_2CO} and \ce{SO_2} are seen to have the same azimuthal extent but the \ce{CH_3OH} is more compact and following the dust, similar to the higher resolution SO observations. 

The total disk integrated line flux was extracted from the channel maps using a Keplerian mask and is listed in Table~\ref{line_list} with a 10\% ALMA flux calibration error. These spectra are included in Appendix Fig~\ref{fig:profiles}. 
There is no evidence for an additional component of emission or broader line-profiles than expected from a disk, as seen in HD~100546 in SO \citep{2018A&A...611A..16B}.

We calculate the average column densities of \ce{SO_2}, \ce{^{34}SO_2} and \ce{SO} (and an upper-limit on the CS column density) from the disk integrated fluxes. For this we assume the line emitting area to be the same as the 5$\sigma$ extent of the 0.9~mm dust emission from \citet{Francis2020}
(see Figure~\ref{fig:dust} in the Appendix). This is the same approach as in Paper I and is an area of 1.2$\times10^{-11}$ steradians.  
Following the method outlined in \citet{1999ApJ...517..209G} and applied to disks \citep[e.g.][]{2018ApJ...859..131L} we construct an opacity-corrected rotational diagram for the \ce{SO_2} resulting in a rotational temperature of $\approx$100~K. This framework assumes LTE and that the lines are optically thin to marginally optically thick. We also calculate column densities for \ce{^{34}SO_2}, \ce{SO} and \ce{CS} assuming an excitation temperature of 50, 100 and 150~K. The latter two are approximately the rotational temperatures derived for \ce{CH_3OH} and \ce{H_2CO} respectively in Paper I. 
The resulting values are listed in Table~\ref{tab:columndens}.
The inferred \ce{SO_2}/\ce{^{32}SO_2} is only 2 compared to the value of 22 for the \ce{^{32}S}/\ce{^{34}S} isotope ratio \citep[e.g.][]{1999RPPh...62..143W}.
This shows that the emission from the primary isotopologue, \ce{SO_2}, is optically thick. 
We therefore do not use the \ce{SO_2} column density derived from the \ce{SO_2} but rather using the rarer isotopologue \ce{^{34}SO_2}. 
The ratio of CS/SO is $<$0.01 and CS/\ce{SO_2} $<$0.02 (using \ce{SO_2} derived from the \ce{^{34}SO_2}).  We note that as the inferred column densities are inversely proportional to the assumed emitting area our results may be a conservative lower limit. 

As a complementary check for optically thick \ce{SO_2} we also calculate the brightness temperature of the \ce{5_{3,3}}-\ce{4_{2,2}} line. This was done by re-imaging the line before continuum subtraction, calculating the peak intensity map (moment 8) and then converting this to a brightness temperature under the Rayleigh-Jeans approximation. We find that this results a low temperature with a peak of $\approx$7~K. This is most likely due to beam dilution as the emission is currently unresolved. A brightness temperature of $\approx$100~K would require an emitting area $\approx$15$\times$ smaller than the 0\farcs55$\times$0\farcs44 beam of these data. Future 0\farcs1 data will be able to test this.

\begin{table}[]
    \centering
    \caption{Derived column densities and ratios}
    \begin{tabular}{cccc}
    \hline \hline
    Molecule & $T_{ex}$ (K) & $\mathrm{N_{T}}$ (cm$^{\mathrm{-2}}$) & $\tau$ \\ \hline
    \ce{SO_2}$^{*}$ & 50 &  2.6 $\pm~0.3$ $\times~10^{15}$ & [$>$ 1]\\
       & 100 &  4.4 $\pm~0.4$ $\times~10^{15}$ & [$>$ 1]\\
   & 150 &  6.6 $\pm~0.4$ $\times~10^{15}$ & [$>$ 1]\\

    \ce{^{34}SO_2} & 50 &  1.2 $\pm~0.1$ $\times~10^{14}$  & [0.26]\\
  & 100 &  2.0 $\pm~0.2$ $\times~10^{14}$ & [0.08]\\
    & 150 &  3.0 $\pm~0.3$ $\times~10^{14}$ & [0.04]\\

    \ce{SO} & 50 & 4.7 $\pm~0.5$ $\times~10^{15}$  & [0.60]  \\ 
  & 100 &  7.1 $\pm~0.7$ $\times~10^{15}$ & [0.16]\\
    & 150 &  9.9 $\pm~1.0$ $\times~10^{15}$ & [0.08]\\
    
\ce{CS} & 50 & $\leq$ 5.9 $\pm$~0.6 $\times~10^{13}$  & [ $\leq$2.3]  \\ 
  & 100 & $\leq$ 2.9 $\pm~0.3$ $\times~10^{13}$ & [$\leq$0.35]\\
  & 150 &  $\leq$ 3.2 $\pm~0.3$ $\times~10^{13}$ & [$\leq$0.14]\\

\hline

\ce{S/H} & 50 & 4.6 $\times~10^{-7}$ & - \\
 & 100 & 7.2 $\times~10^{-7}$  & - \\
 & 150 &  1.0 $\times~10^{-6}$  & - \\

\ce{CS/SO} & 50 & $\leq$0.012 & - \\
 & 100 & $\leq$0.004 & - \\
 & 150 & $\leq$0.003 & - \\
 
 \ce{SO/SO_2} & 50 & 1.8  & - \\
 & 100 & 1.6  & - \\
 & 150 &  1.5 & - \\

   \hline
    \end{tabular}
    \begin{tablenotes}\footnotesize
\item{$^{*}$ using \ce{^{34}SO_2} and \ce{^{32}S}/\ce{^{34}S} = 22}
\end{tablenotes}
    \label{tab:columndens}
\end{table}

\section{Discussion}

\subsection{Volatile sulphur in the IRS 48 disk}

The bulk of the elemental sulphur in protoplanetary disks is thought to be in refractory materials ($\approx$89$\pm$8$\%$) \citep{2019ApJ...885..114K}. Therefore, it is interesting to calculate what fraction of the expected volatile sulphur abundance in the IRS~48 disk can be accounted for by the \ce{SO} and \ce{SO_2} we detect. To estimate this we take the total volatile sulphur to be the sum of the \ce{SO_2} (from the \ce{^{34}SO_{2}}) and \ce{SO} average column densities. The ${N_{\mathrm{H}}}$ column density at 60~au from the IRS~48 disk model is $1.6\times10^{22}$~$\mathrm{cm^{-2}}$ where the gas density structure is constrained from \ce{C^{17}O} observations \citep{Bruderer2014, vanderMarel2016-isot}. This results in a S/H ratio of 4.6-10.0$\times10^{-7}$ (depending on the excitation temperature, see Table~\ref{tab:columndens}) which is $\approx$15-100$\%$ of the total range in volatile S budget estimated by \citet{2019ApJ...885..114K}. Although we can account for the bulk of the volatile sulphur in the disk with just the \ce{SO} and \ce{SO2} molecules, further observations targeting \ce{H_2S} and both oxygen and carbon bearing sulphur molecules (\ce{H_2CS}, \ce{CH_3SH}, \ce{OCS} and \ce{C_2S}) will provide the complete picture as to how the volatile sulphur is partitioned in this disk.

\subsection{Chemical origin of the SO and \ce{SO_2} in the IRS~48 disk}

We find that \ce{SO} and \ce{SO_2} gas in the IRS~48 disk is co-spatial with the asymmetric dust trap. 
If the \ce{SO} and \ce{SO_2} are forming in the gas phase then the 
key reactions are \citep[e.g.][]{1997ApJ...481..396C}:
\begin{equation*}
    \ce{S} + \ce{OH} \rightarrow \ce{SO} + \ce{H},
\end{equation*}
\begin{equation*}
    \ce{SO} + \ce{OH} \rightarrow \ce{SO_2} + \ce{H}. 
\end{equation*}
Therefore gas phase formation of \ce{SO} and \ce{SO_2} requires sufficient OH in the disk atmosphere. Interestingly \textit{Herschel}/PACS reported non-detections of both OH and \ce{H_2O} in the IRS-48 disk \citep{2013A&A...559A..77F}. Additionally as the molecules are co-spatial with the dust trap this indicates that gas-grain processes are important to consider too. The alternative origin for the gas-phase \ce{SO} and \ce{SO_2} that we propose is from the sublimation of ices, particular at the dust cavity wall \citep{2011ApJ...743L...2C, booth2021}. If the ice reservoir is constrained to the larger millimeter and cm-sized dust grains, as suggested for TW~Hya \citep{2016A&A...591A.122S, 2016ApJ...823L..10W}, then this would explain the morphology of the molecular emission we see towards IRS~48. This could be either from the direct sublimation of \ce{SO} and \ce{SO_2} ice or via gas-phase chemistry following the UV photo-dissociation of evaporated \ce{H_2S} and \ce{H_2O} ices. 
The sublimation temperatures of \ce{SO_2} and \ce{H_2CO} are similar so the azimuthal correlation of these species makes sense, and it follows that the \ce{CH_3OH} is more compact. As the SO data are of much higher spatial resolution and lower signal-to-noise a direct comparison to the other line spatial morphologies is not too meaningful. 

The vortex in the IRS~48 disk may also play a role in the chemistry.
Vertical convection cells in the vortex might also transport icy material from the disk midplane up into the warm molecular layer resulting in an increase in the sublimation of ices \citep{2010A&A...516A..31M} as suggested also for \ce{H_2CO} and \ce{CH_3OH} (see Figure 4 in Paper I). In line with our results exploratory chemical models from \citet{2020IAUS..345..285D} show that SO is the primary S-carrier in the gas phase in a IRS~48 like disk model. They also show that the \ce{SO_2} abundance is sensitive to the dust-to-gas mass ratio in the disk, therefore the SO/\ce{SO_2} ratio may be a useful complementary tracer of the gas and dust properties of vortices. 

A particular disk to compare to IRS~48 is AB~Aur as they both host azimuthal dust traps. In AB~Aur SO has been detected and shows a ring-like morphology \citep{2020A&A...642A..32R}. SO emission over the full azimuth could be explained by AB~Aur hosting a less azimuthally concentrated dust trap than in IRS~48 as the millimeter grains are detected throughout the whole azimuth of the AB~Aur disk \citep[see Figure 2 in][, note this is not Papper I, for a comparison of the mm-dust emission from both disks]{2021AJ....161...33V}. Therefore, this is still consistent with our proposed scenario for the IRS~48 disk hosting a particularly massive ice trap with a low gas to dust ratio at the trap. 

We derive a \ce{SO}/\ce{SO_2} ratio of 1.5-1.8 (see Table~\ref{tab:columndens}) which is closest with the values found for Hale-Bopp \citep[\ce{SO}/\ce{SO_2}=1.3;][]{2000A&A...353.1101B} and the Class I disk Elias~29 \citep[\ce{SO}/\ce{SO_2}=2.0;][]{2019ApJ...881..112O}. 
These values are a factor of a few higher than measured in 
IRAS~16293-2422 B (\ce{SO}/\ce{SO_2}=0.3) and the comet 67P (\ce{SO}/\ce{SO_2}=0.4-0.7) \citep{2016MNRAS.462S.253C,2018MNRAS.476.4949D}. 
The chemistry in the disk is not solely due to thermal desorption but is also driven by the UV and X-ray irradiation fields. This would result in the photo-dissociation of \ce{SO_2} to form \ce{SO} accounting for the higher SO/\ce{SO_2} ratio we observe towards IRS 48 than IRAS~16293-2422 B. 
In photo-dissociation regions this value is indeed higher with, e.g.  \ce{SO}/\ce{SO_2}$\approx$10 in the Horsehead Nebula \citep{2019A&A...628A..16R} and $\approx$3 in the extended ridge of the Orion Bar compared with 0.4 at the location of the hot core \citep{2014ApJ...787..112C}.


\subsection{Constraints on the C/O ratio of the gas}

The non-detection of CS in a disk that is rich in emission from volatile sulphur molecules may appear puzzling. Our 3$\sigma$ upper-limit on the CS column density is at the high end of the values detected in other sources which cover more than one order of magnitude ($\times10^{12}-10^{13}$ cm$^{-2}$) \citep[][]{2018A&A...617A..28S, 2019ApJ...876...72L}. 
In these other disks the inferred CS/SO ratios are $>$1000$\times$ higher than our observations (CS/SO $<$ 0.01)
and the detection of CS and the non-detection of SO (and \ce{SO_2}) have been used to infer the C/O ratio in the warm molecular layer. Chemical modelling shows that the CS/SO abundance ratio is the most sensitive to the overall C/O ratio in the gas rather than other physical/chemical processes, e.g., grain growth, UV irradiation and turbulent mixing \citep{2018A&A...617A..28S,2020A&A...638A.110F}.
Disk specific modelling efforts show that CS/SO ratios of order 100 
can be re-produced via gas-phase chemistry where the C/O ratio is super-solar ($>$1) \citep[][]{2011A&A...535A.104D, 2018A&A...617A..28S}. 
Even in AB~Aur, where SO is detected, an elevated C/O ratio of 1.0 is favoured in the chemical modelling \citep{2020A&A...642A..32R}. Recent observations of the transition disk PDS~70 also report a CS/SO $>$100 which is in line with a C/O$>$1 \citep{2021arXiv210108369F}.  This means that the bulk of gas accreted by forming planets in these disks is oxygen poor. 

In the IRS~48 disk we have the opposite situation where the CS/SO $<$0.01 (see Table~\ref{tab:columndens}). From a comparison to chemical models our column density ratios suggest a solar C/O ratio \citep{2018A&A...617A..28S}.  A complementary test of high C/O ratio in IRS 48 would be the detection/non-detection of strong \ce{C_2H} emission \citep[e.g.][]{2016ApJ...831..101B} and other oxygen rich molecules like \ce{H_2O}. 
In general, the sublimation of ices at the edge of the cavity means that the gas accreted through the cavity onto potential forming planets will have significantly different elemental composition to the gas accreted by a planet forming in a gap further out in a disk.
This is in contrast to what has been proposed for TW~Hya \citep{2019A&A...632L..10B} where the dust trap is beyond the CO ice line and thus the inner disk gas is volatile poor. 

Another transition disk with an azimuthal asymmetry is HD~142527. CS has been detected in this disk and is also asymmetric but it is anti-coincident with the dust trap peaking on the other side of the disk \citep{2014ApJ...792L..25V}. The authors propose that this is either due to the lower temperatures in the dust trap resulting in the freeze-out of molecules or the high optical depth of the dust obstructing the line emission from the dust trap region. But, it could alternatively be tracing an azimuthal change in the C/O ratio where C/O $<$1 in the region of the dust trap due to the sublimation of ices.

\section{Conclusions}

In this letter we have presented ALMA observations targeting the S-bearing molecules \ce{SO}, \ce{SO_2} and \ce{CS} in the unique Oph-IRS~48 disk hosting a massive ice trap. We summarise our conclusions here:

\begin{itemize}
    \item This is the first robust detection of \ce{SO_2} and the isotopologue \ce{^{34}SO_2} in a Class II disk.
    \item These oxygen rich S-bearing molecules have a clear disk origin unlike in other sources where they trace shocks or winds/outflows.
    \item As the sulphur species are co-spatial with the dust trap we suggest that presence of the gas-phase \ce{SO} and \ce{SO_2} is due to the sublimation of ices at the edge of the dust cavity and that the bulk of the ice reservoir is coincident with the millimeter, and larger, dust grains. 
    \item We estimate the average S/H from the \ce{SO} and \ce{SO_2} column densities to be $4.6-10.0\times10^{-7}$ and this is consistent with 15-100~\% of the total expected volatile sulphur budget in the disk. 
    \item Unlike all other protoplantary disks targeted in these molecules so far, CS is not the primary S-carrier in the gas phase and the CS/SO ratio is $<$0.01 in strong contrast to other disks.
    \item The inferred gas-phase C/O ratio $<$1 and is likely solar. We propose that this is due to the chemistry being driven by the sublimation of ices rather than gas-phase processes. This is a result of the unique nature of the extreme dust trap in the IRS~48 disk. This has important implications for the elemental composition of the atmosphere of gas-giant planets that maybe be forming or will form in this disk.  
\end{itemize}

This work and Paper I further showcases the use of transition disks as unique observational laboratories to explore the full volatile inventory in disks due to the sublimation of ices at the cavity edge. 

\bibliographystyle{aa} 
\bibliography{myrefs} 

\begin{acknowledgements}
We would like to thank Akimasa Kataoka for his help with the reduction of the
data. N.M. acknowledges support from the Banting Postdoctoral
Fellowships program, administered by the Government of Canada. ALMA is
a partnership of ESO (representing its member states), NSF (USA) and NINS
(Japan), together with NRC (Canada) and NSC and ASIAA (Taiwan) and KASI
(Republic of Korea), in cooperation with the Republic of Chile. The Joint ALMA
Observatory is operated by ESO, AUI/ NRAO and NAOJ. This paper makes use
of the following ALMA data: 2013.1.00100.S, 2017.1.00834.S.
\end{acknowledgements}

\newpage

\begin{appendix} 

\section{Additional Figures}

\begin{figure*}
    \centering
    \includegraphics[width=\hsize]{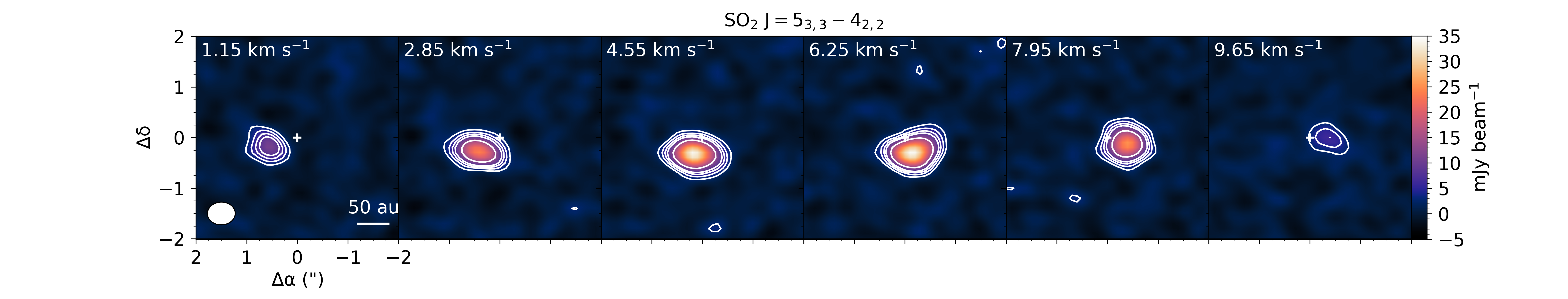}
    \includegraphics[width=\hsize]{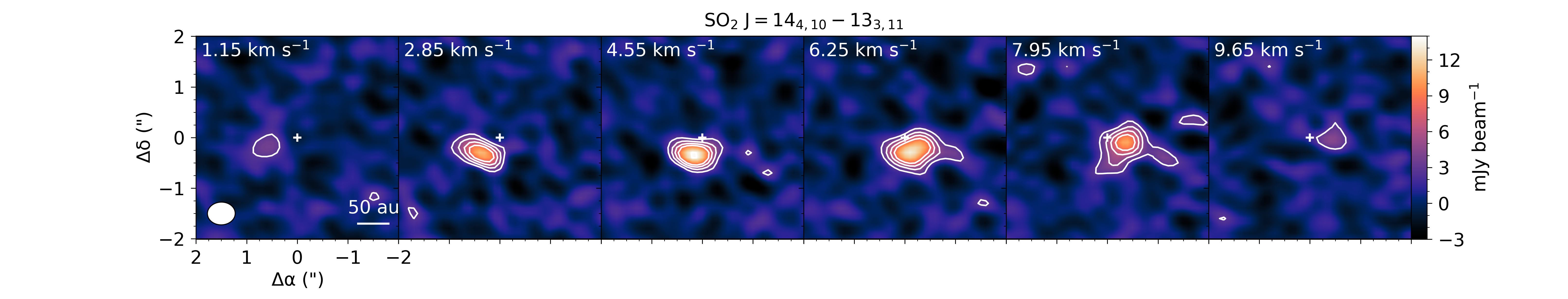}
    \includegraphics[width=\hsize]{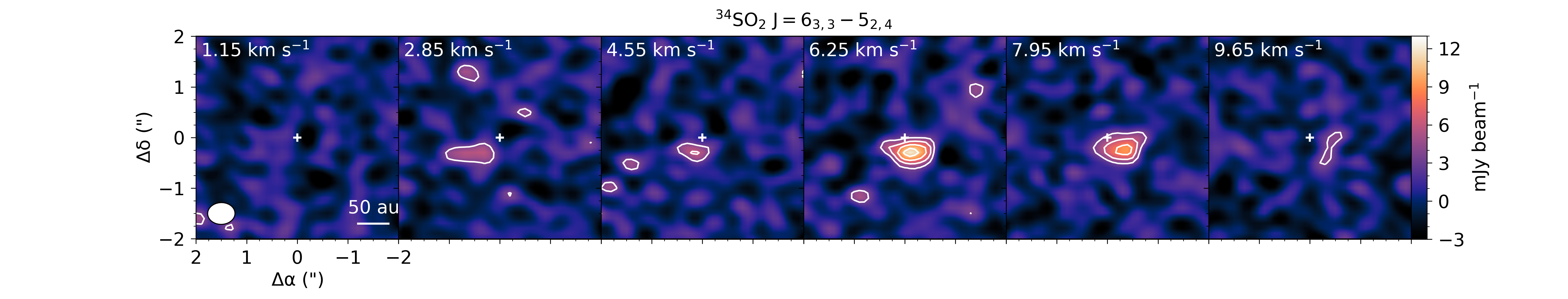}
    \includegraphics[width=\hsize]{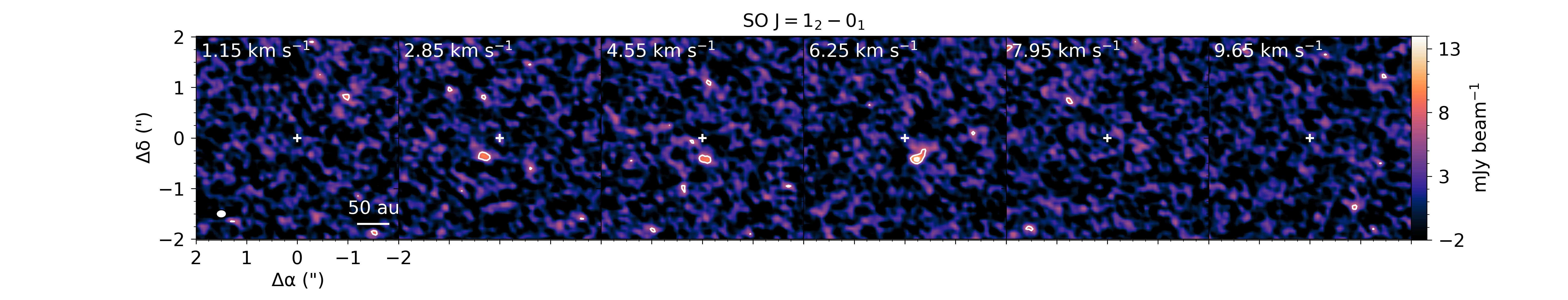}
    \caption{Channel maps of the detected \ce{SO}, \ce{SO_2} and \ce{^{34}SO_2} where the contours mark [3, 5, 7, 9, 15]$\times~\sigma$ where $\sigma$ is the rms listed in Table~\ref{line_list}.}
    \label{fig:channelmaps}
\end{figure*}

\begin{figure*}
    \centering
    \includegraphics[width=\hsize]{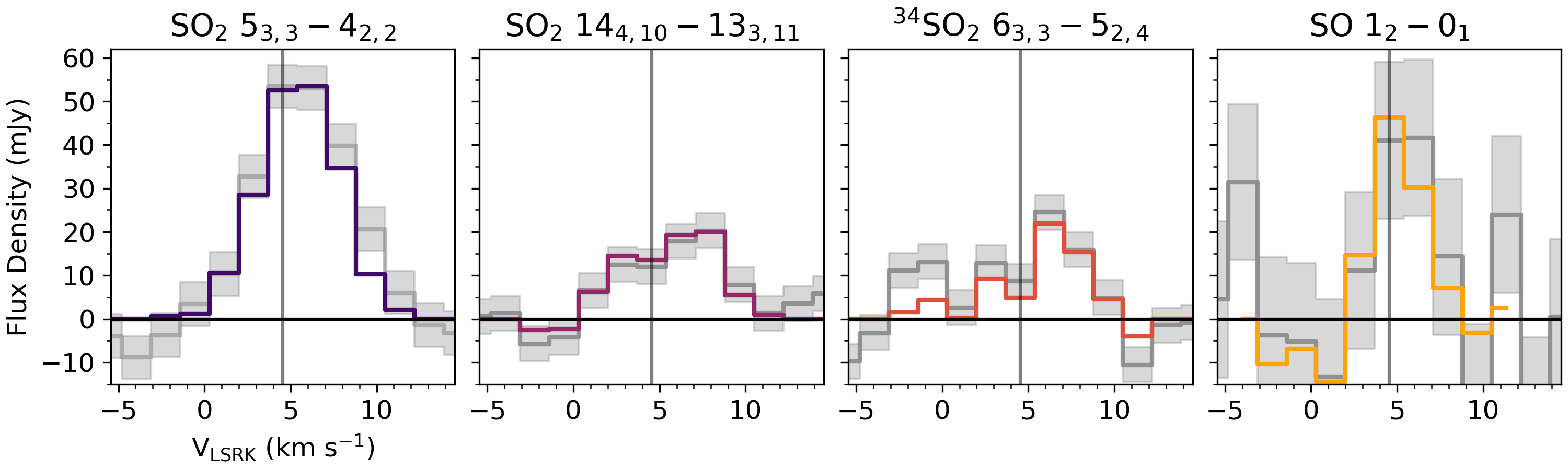}
        \includegraphics[width=\hsize]{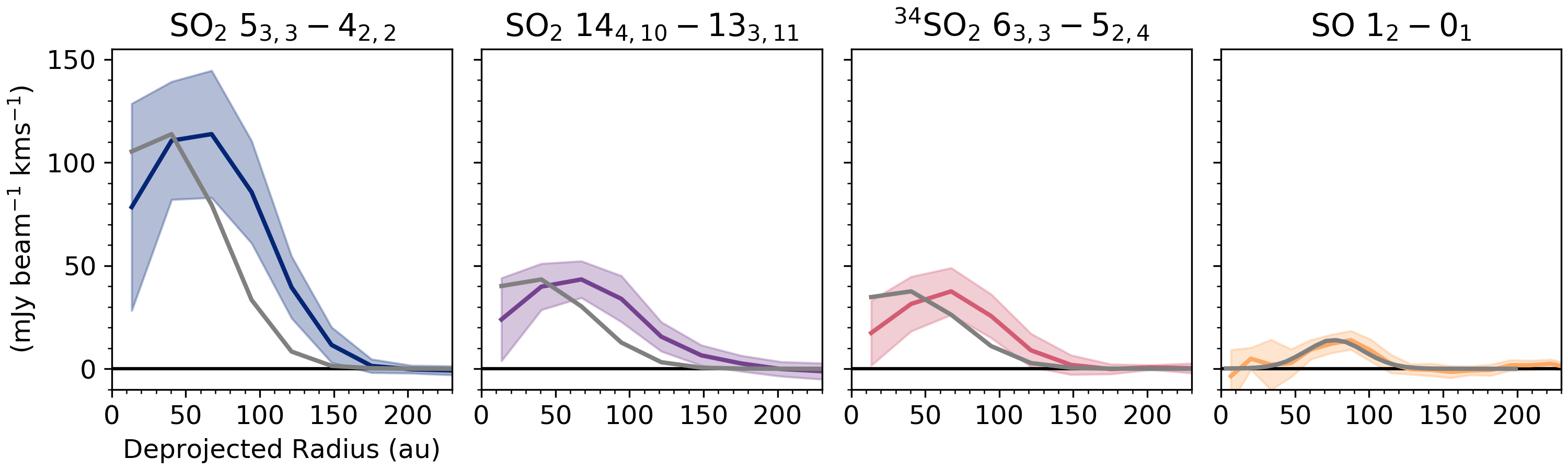}
        \includegraphics[width=\hsize]{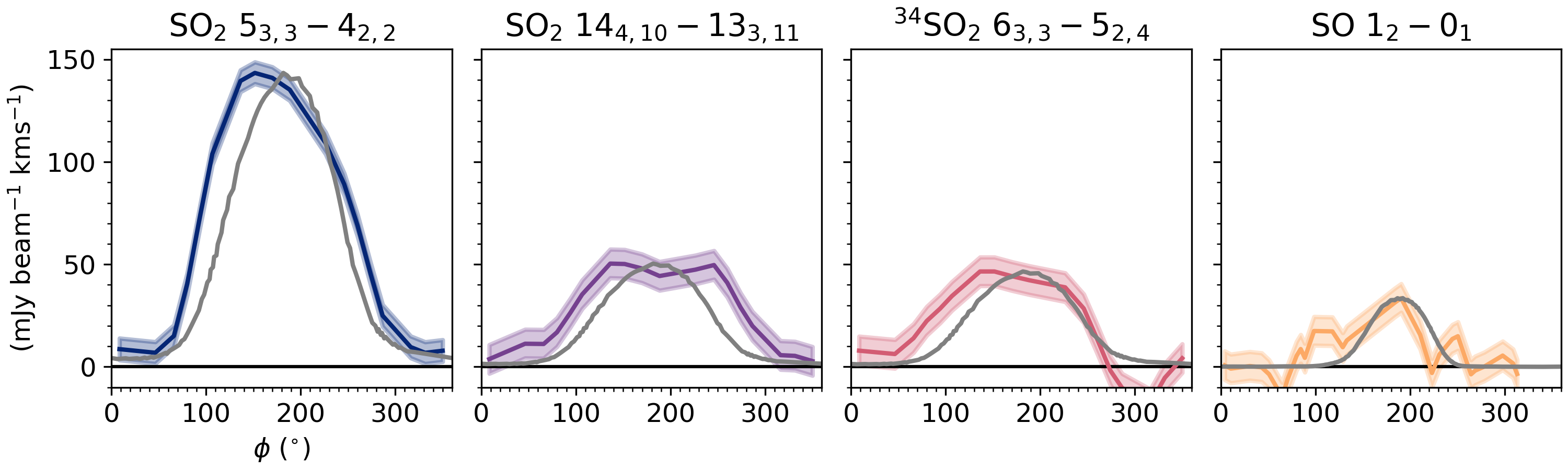}
    \caption{Top: Spectra extracted from Keplerian masks (colours) and 1\farcs5 ellipse centred on source (grey) with $\pm$1~$\sigma$ error bars. Grey line marks the source velocity (4.55~km~$\mathrm{s^{-1}}$).
    Middle: Radial emission profiles averaged over wedge from $100^{\circ}$ to $260^{\circ}$. Grey profiles are the dust emission at the same resolution as the line data and error bars on the line profiles are shaded regions. Bottom: Azimuthal profiles extracted at a deprojected radius of 70~au for the \ce{SO_2} lines and 90~au for the SO lines. The grey lines show the dust azimuthal profile extracted at the same radius for each line respectively.}
    \label{fig:profiles}
\end{figure*}


\begin{figure}
    \centering
    \includegraphics[width=\hsize]{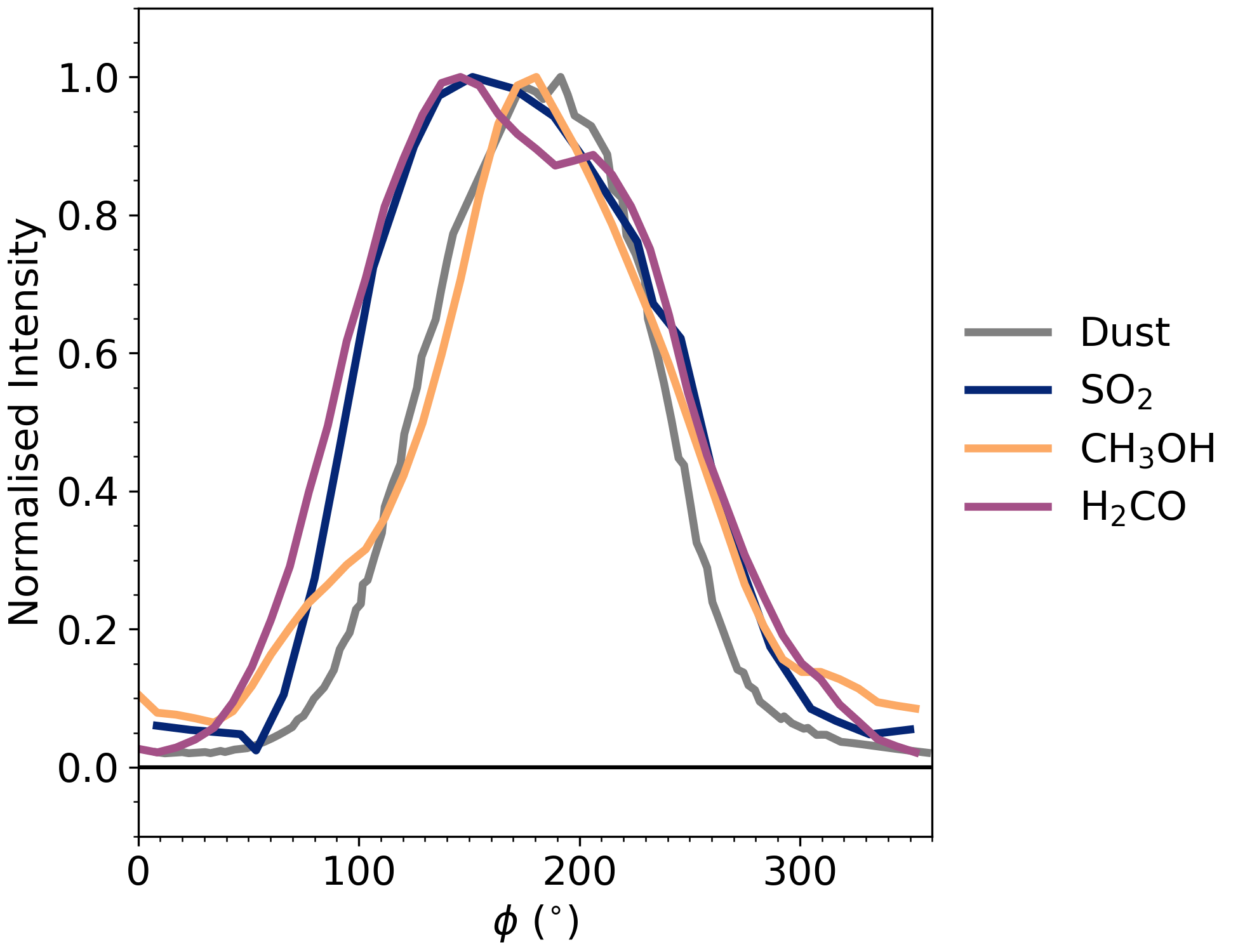}
    \caption{Normalised azimuthal profiles of 0.9~mm dust, \ce{H_2CO} (transition), \ce{CH_3OH} (transition) and \ce{SO_2} ($J=5_{3,3}-4_{2,2}$) lines taken at 70~au. The \ce{H_2CO} and \ce{CH_3OH} data are from \citet{vanderMarel2021}.}
    \label{fig:comparison}
\end{figure}


\begin{figure}
    \centering
    \includegraphics[width=\hsize]{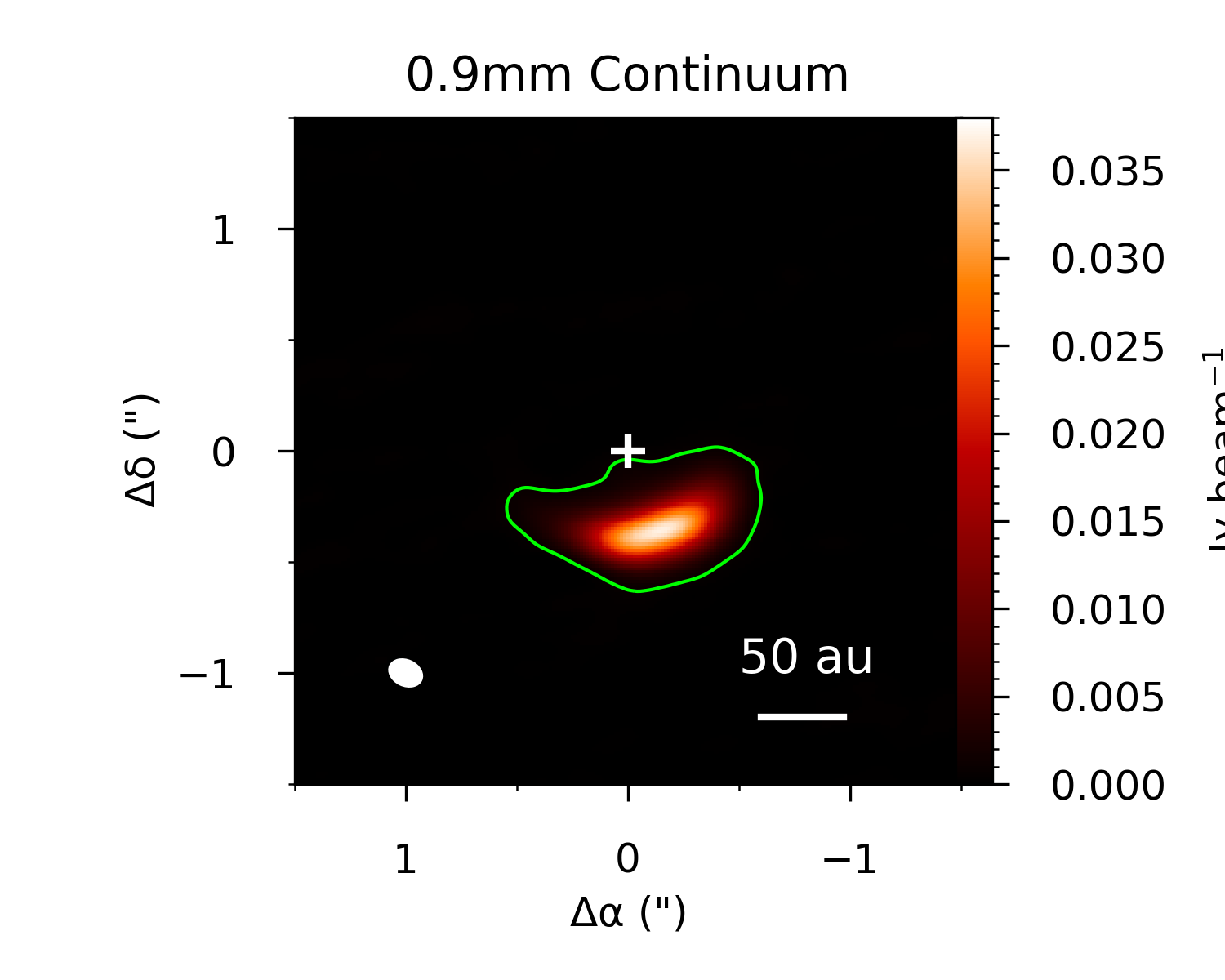}
    \caption{0.9~mm continuum emission from \citet{Francis2020}. Green contour is the 5~$\sigma$ level and is the area used to calculate the column densities.}
    \label{fig:dust}
\end{figure}

\end{appendix}
\end{document}